\documentclass[
 aip,cha,
 amsmath,amssymb,
 reprint,
]{revtex4-2}

\usepackage{graphicx}
\usepackage{dcolumn}
\usepackage{bm}
\usepackage[caption=false]{subfig}
\usepackage[utf8]{inputenc}
\usepackage[T1]{fontenc}
\usepackage{mathptmx}
\usepackage{etoolbox}
\usepackage{amsmath}
\usepackage{mathtools}
\usepackage[monochrome]{xcolor}
\newcommand{\D}[0]{{\rm d}}
\newcommand{\ie}[0]{\textit{i.e.}}
\newcommand{\img}[0]{{\rm j}}
\newcommand{\etal}[0]{\textit{et~al.}}
\usepackage{hyperref}
\hypersetup{
    colorlinks = true,
    citecolor = blue,
    linkcolor = red,
}

\makeatletter
\def\@email#1#2{%
 \endgroup
 \patchcmd{\titleblock@produce}
  {\frontmatter@RRAPformat}
  {\frontmatter@RRAPformat{\produce@RRAP{*#1\href{mailto:#2}{#2}}}\frontmatter@RRAPformat}
  {}{}
}%
\makeatother
\begin{document}

\preprint{AIP/123-QED}

\title[On the role of higher-order interactions towards first synchronization time]{On the role of higher-order interactions towards first synchronization time}

\author{Dhrubajyoti Biswas}
\email{dhrubajyoti98@gmail.com; dhrubajyoti@phy.iitkgp.ac.in; Corresponding Author}
\affiliation{Indian Institute of Technology Kharagpur, West Bengal, 721302, India}

\author{Pintu Patra}
\email{pintupatra@phy.iitkgp.ac.in}
\affiliation{Indian Institute of Technology Kharagpur, West Bengal, 721302, India}

\author{Arpan Banerjee}%
\email{arpan@nbrc.ac.in, banerjee2007@gmail.com}
\affiliation{National Brain Research Centre, Manesar, Gurgaon, Haryana, 122052, India}

\date{\today}

\begin{abstract}
\textcolor{red}{This study investigates transient collective dynamics, with a focus on how higher-order interactions impact the time required to reach steady-state synchronization.} Assuming a large ensemble of \textcolor{red}{deterministic} and globally coupled Kuramoto oscillators with Cauchy-distributed natural frequencies, an expression for the first synchronization time is derived using the Ott-Antonsen ansatz. Subsequent numerics reveal that (i) increasing the coupling strengths for a fixed interaction order accelerates the transition to synchronization and (ii) increasing the interaction order for fixed interaction strength produces non-monotonic behavior. In particular, the inclusion of triadic interactions generally accelerates synchronization, whereas further higher-order interactions progressively delay convergence to the steady state, in some regimes even falling below the pairwise level. Ultimately, for very large interaction orders, the dynamics revert to pairwise-like behavior. Simulations of the system equations for different parameter combinations support these observations, while the asymptotic case is interpreted through the nonlinear structure of the order-parameter dynamics.
\end{abstract}

\maketitle

\begin{quotation}
Higher-order interactions are crucial components of real physical systems. However, while their role in shaping steady-state dynamics is well studied, their influence on transient dynamics remains less understood. This work investigates one such transient measure, namely, the first synchronization time in a globally coupled Kuramoto model, and its dependence on the strength and order of interaction. The obtained results indicate that, although increasing the coupling strength monotonically speeds up synchronization, increasing the interaction order yields a non-monotonic response that can be attributed to the nonlinear structure of the order parameter dynamics.
\end{quotation}

\section{Introduction}
\label{intro}

\begin{figure}[t]
    \centering
    \includegraphics[width=\linewidth]{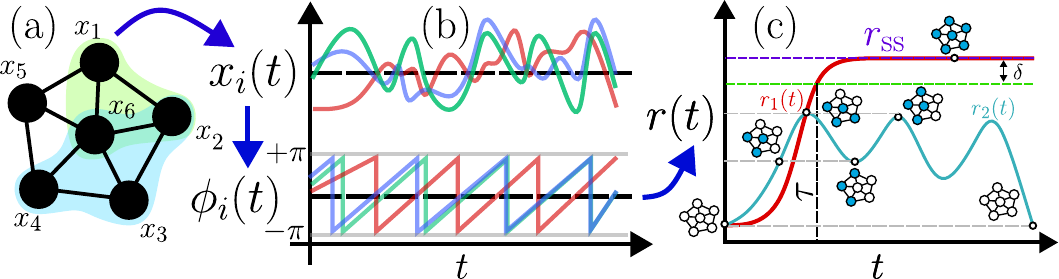}
    \caption{(a): Network schematic showing pairwise and higher-order (here, triadic and quartic) interactions, via black lines and coloured regions, respectively; (b):  Temporal evolution of the state variables $x_i(t)$ (top), and their corresponding phases $\phi_i(t)$ (bottom), obtained through appropriate approximations;  (c): Schematic diagram of the global order parameter $r(t)$, showing two qualitatively different realizations. Here, $r_1(t)$ (red) reaches a steady-state at $r_{\rm ss}$ (horizontal magenta line), whereas $r_2(t)$ (blue) does not. The horizontal green line indicates the boundary of the region $[r_{\rm ss}-\delta,r_{\rm ss}]$, whereas the vertical black line at $t=\tau$ denotes the first synchronization time. The configurations at different instances highlight the system's state, where the subset of nodes in blue indicates high synchronicity.}
    \label{demo}
\end{figure}

Emergent phenomena in physical systems arise due to the interplay between the structure and the nodal dynamics of the underlying network. However, while a significant amount of literature has focused on pairwise interactions, recent studies~\cite{majhi2022dynamics,battiston2020networks} highlight the importance of interactions that encapsulate three or more nodes simultaneously; see Fig.~\ref{demo}a for a toy schematic. These interactions form a separate class, termed ``higher-order interactions'', and are motivated by examples ranging from social networks~\cite{wang2020social} to chemical reactions~\cite{marehalli2025hypergraph}, and are particularly prevalent in biological systems~\cite{grilli2017higher,tekin2017measuring,mayfield2017higher,letten2019mechanistic} such as the brain~\cite{martignon1995detecting,yu2011higher,huang2017weak,sizemore2018cliques,karmelic2022emergent,PhysRevResearch.3.043193}. On the other hand, synchronization is an ubiquitous emergent phenomenon~\cite{PhysRevE.98.032217,arenas2008synchronization,wu2024synchronization,ghosh2022synchronized} observed in such systems, often characterized by near-identical temporal evolution of the variables that represent the system’s instantaneous state. It includes examples that range from the collective flashing of fireflies in a jungle to coordinated neural activity in the human brain~\cite{sarfati2020spatio,sarfati2023emergent,henao2020entrainment,hovel2020synchronization,ranjan2024propagation}, understanding the mechanisms of which are of practical importance to several disciplines. This prompts active research into methods of controlling synchronization, where the Kuramoto model of coupled phase oscillators~\cite{RevModPhys.77.137} has emerged as a simple and paradigmatic framework in the study of synchronization, owing to its tractability and broad applicability across domains~\cite{gupta2014kuramoto,bayani2023explosive,crnkic2021synchronization,frolov2021extreme,PhysRevX.9.011002,asano2025synthesized,PhysRevB.105.174305,ameli2024two,ameli2021time,jin2023synchronization,biswas2024effect,o2017oscillators,PhysRevE.105.014211,o2025global,pathak2022biophysical}. Importantly, a broad range of complex oscillatory dynamics can be typically mapped onto the Kuramoto framework through certain approximations~\cite{banerjee2007neural,pillai2017symmetry}; see Fig.~\ref{demo}b.

There has also been considerable focus on exploring the properties of higher-order interactions and their interplay with collective phenomena. Some recent and key contributions in this direction include Millan~\etal~\cite{millan2020explosive}, where the authors introduce a novel formulation by allowing the state variables to reside on higher-dimensional structures, rather than only on nodes; Dai~\etal~\cite{dai2025higher}, where the authors investigate a three-oscillator Kuramoto model with higher-order couplings and demonstrate that different functional forms generate a rich repertoire of collective patterns including partial synchronization and abrupt desynchronization; and Xiang~\etal~\cite{xiang2026synchronization}, where the authors show that excessively strong higher-order interactions promote clustered states. Furthermore, while these studies generally investigate the forward problem, \ie, they determine the dynamical outcomes for a given network structure and coupling function, there is also growing interest in the inverse problem of uncovering the underlying higher-order network structure from empirical and synthetic data, with studies such as those by Tabar~\etal~\cite{PhysRevX.14.011050} and Malizia~\etal~\cite{malizia2024reconstructing} being some of the recent important contributions. However, despite these and other such efforts~\cite{muolo2025higher,berte2025fibration,zqf8-tg6g}, the added complexity of higher-order interactions implies that their overall effects remain to be fully characterized, especially for interactions that go beyond triadic and quartic order.

While most of the existing studies focus primarily on the steady-state synchronous dynamics, the transient en route to the steady-state is also of considerable importance in many physical settings. This is because the steady-state measures alone do not fully characterize the collective dynamics of complex systems, and systems exhibiting similar stationary properties may demonstrate substantially different transient dynamics, indicative of some important underlying differences. In this context, a study by Sinha~\etal~\cite{sinha2023statistics} established the first synchronization time (abbreviated as FST and precisely defined later) as a quantifiable measure of transient synchronization dynamics and showed that its distribution follows a Gumbel shape for the pairwise Kuramoto model. On a similar note, Zhao~\etal~\cite{zhao2026synchronization} studied a triadic Kuramoto model with non-Gaussian noise and highlighted intermittent coherent states and nontrivial first-passage dynamics, whereas Ham~\etal~\cite{ham2024stochastic} demonstrated the role of noise on the first passage time in cellular events, further affirming that transient dynamics is a key dynamical metric across disciplines, and higher-order interactions could play an important role in determining its properties. Building upon these, the current work assumes a globally coupled model of Kuramoto phase oscillators, akin to Sinha~\etal~\cite{sinha2023statistics} and Zhao~\etal~\cite{zhao2026synchronization}, while extending the form of the interaction function beyond pairwise and triadic couplings. The interaction order serves as a tunable parameter, thereby enabling a systematic investigation of how higher-order interactions influence transient dynamics.

The rest of this paper is arranged as follows -- Section \ref{mr} introduces the governing equations for the system under consideration and obtains the analytical expression for the first synchronization time. Section \ref{num} discusses the various numerical observations that include the validation of analytical results (Section \ref{validation}), followed by the effects of varying the coupling strengths for a fixed order of interaction, as well as the overall order of interaction (Section \ref{var-results}). The paper ends with a summary of results and future directions in Section \ref{conc}.

\section{Mathematical Model}
\label{mr}

To capture the effects of higher-order interactions, the globally-coupled and \textcolor{red}{deterministic} Kuramoto model is modified as~\cite{biswas2026emergent}
\begin{align}
\begin{split}
    \frac{\D \phi_i(t)}{\D t}=\omega_i &+ \frac{\epsilon_1}{N}\sum_{j_1=1}^N\sin(\phi_{j_1}-\phi_i)\\
    &+\frac{\epsilon_d}{N^d}\sum_{j_1,\dots,j_d=1}^N\sin\left(d\phi_{j_d}-\sum_{k=1}^{d-1}\phi_{j_k}-\phi_i\right),
\end{split}
\label{main-eqn}
\end{align}
where $\phi_i(t)$ and $\omega_i$ denote the instantaneous phase and natural frequency, respectively, of the $i^{\rm th}$ oscillator within an ensemble of size $N$.
Here, $\epsilon_1$ and $\epsilon_d$ quantify the strength of the pairwise and $d$-order interactions, respectively, where the functional forms of the interactions are chosen such that it remains invariant under rotation. The values of $\omega_i$ are sampled from a unimodal distribution, typically chosen to be a Cauchy distribution since it permits an exact low-dimensional reduction in the Ott-Antonsen framework~\cite{ott2008low,ott2009long}. However, a similar low-dimensional approximation remains possible even for other distributions~\cite{campa2022study}. 
\begin{figure}[b]
    \centering
    \subfloat[$\epsilon_d=0$\label{fig_r_ts_e1_6a}]{\includegraphics[width=0.495\linewidth]{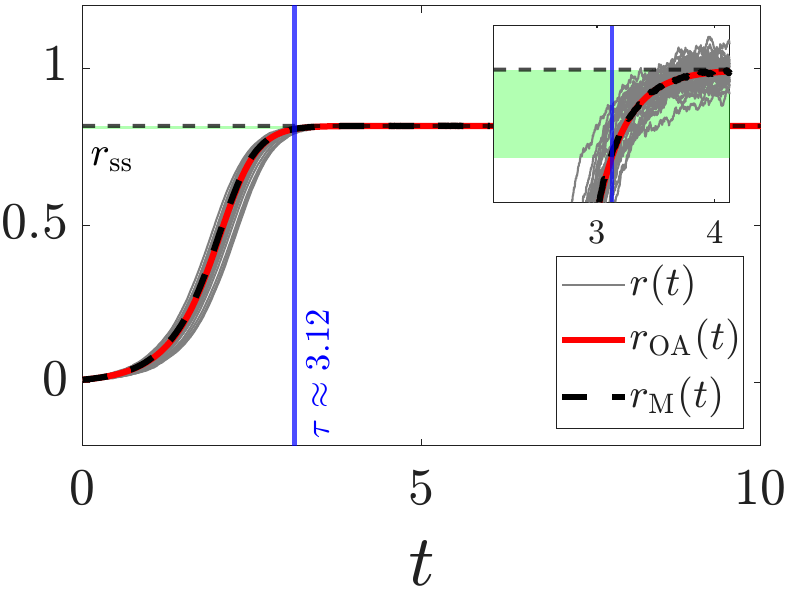}}
    \subfloat[$d=2,\epsilon_d=1$\label{fig_r_ts_e1_6b}]{\includegraphics[width=0.495\linewidth]{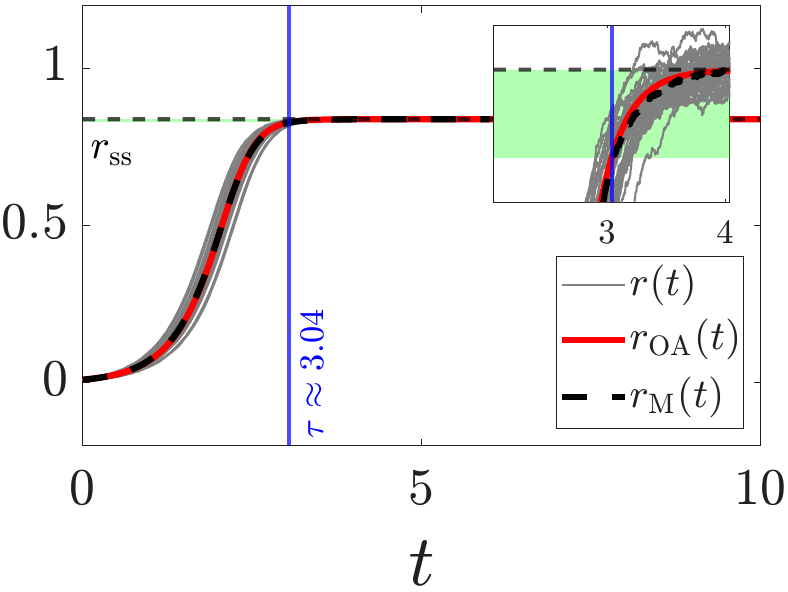}}
    
    \subfloat[$d=3,\epsilon_d=5$\label{fig_r_ts_e1_6c}]{\includegraphics[width=0.495\linewidth]{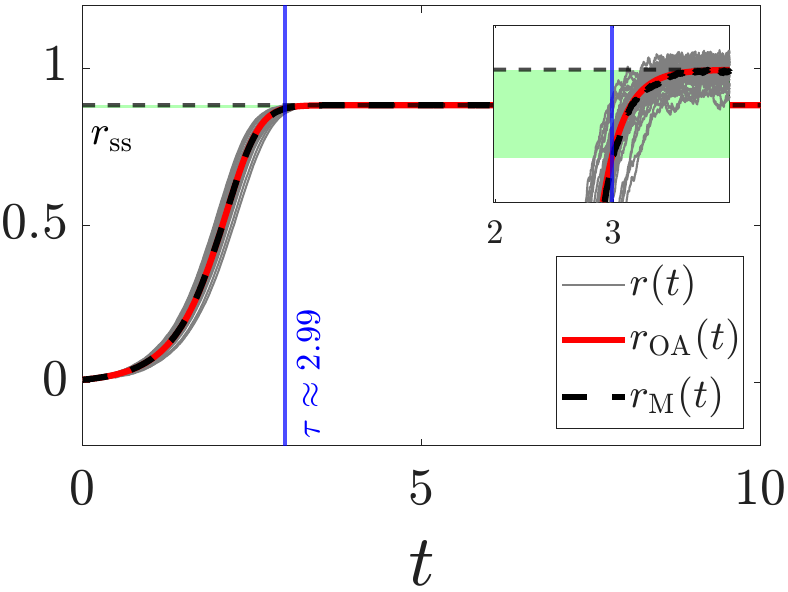}}
    \subfloat[$d=4,\epsilon_d=9$\label{fig_r_ts_e1_6d}]{\includegraphics[width=0.495\linewidth]{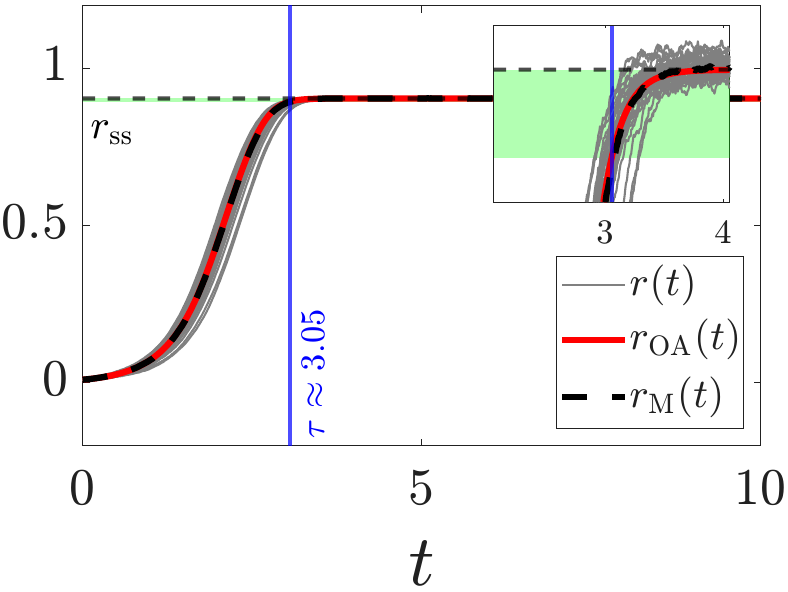}}
    \caption{(a)-(d): Variation of $r$ (in gray), $r_{\rm OA}$ (in red), and $r_{\rm M}$ (in black) as a function of $t\in[0,10]$, for $\epsilon_1=6$ and different values of $(d,\epsilon_d)$. Here, the broken horizontal line demarcates the value of $r_{\rm ss}$, the vertical line denotes the numerically computed value of $\tau$, and the region marked in green denotes the interval $[r_{\rm ss}-\delta,r_{\rm ss}]$. The inset provides a magnified view of the plot where the system’s trajectory moves into the predefined neighborhood of the steady state.}
    \label{fig_r_ts_e1_6}
\end{figure}

\begin{figure*}[t]
    \centering
    \subfloat[\label{e1_vary}]{\includegraphics[width=0.2\linewidth]{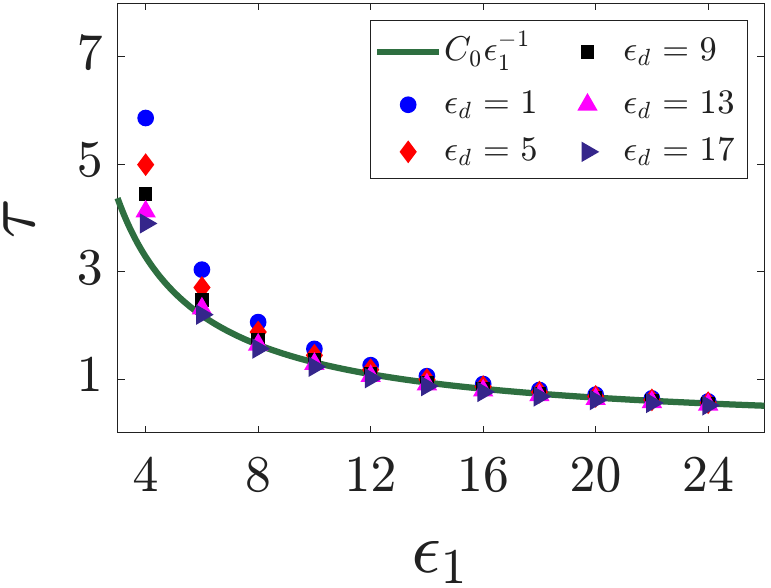}}
    \subfloat[\label{ed_vary}]{\includegraphics[width=0.2\linewidth]{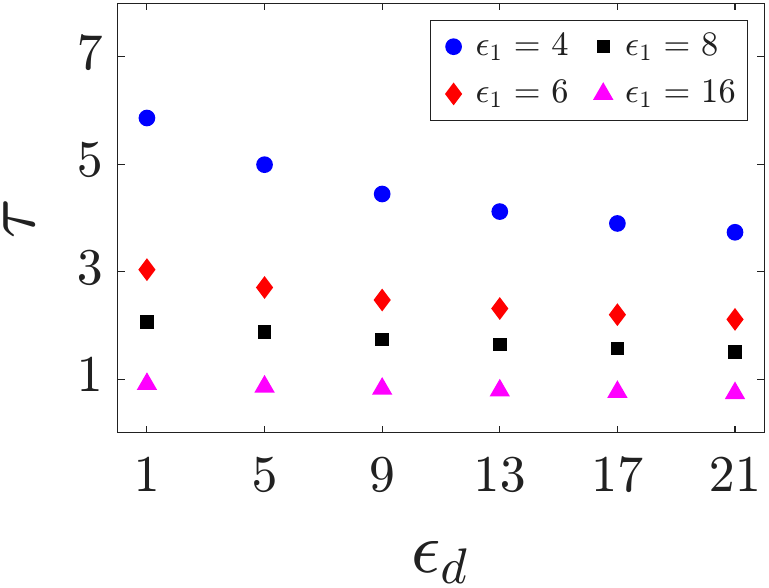}}
    \subfloat[$\epsilon_d=1$\label{rts_ed_1_d_2}]{\includegraphics[width=0.2\linewidth]{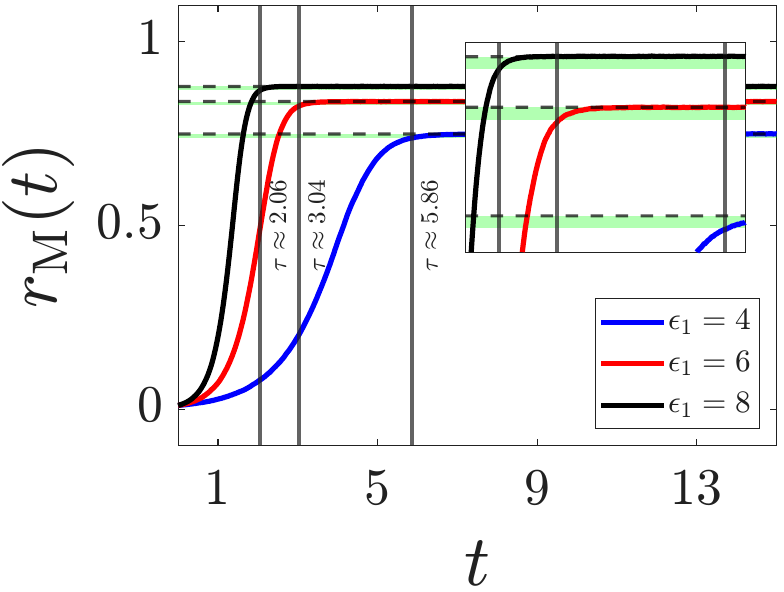}}
    \subfloat[$\epsilon_1=6$\label{rts_e1_6_d_2}]{\includegraphics[width=0.2\linewidth]{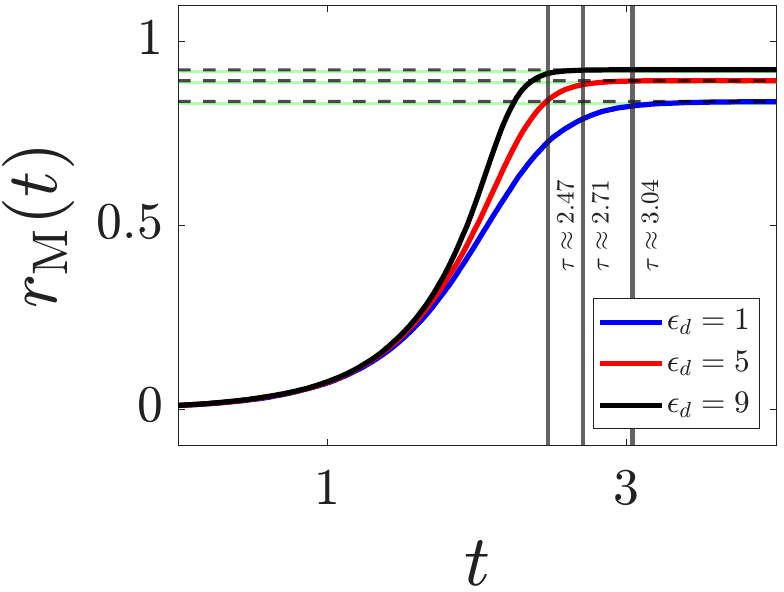}}
    \subfloat[\label{param_plot_d2}]{\includegraphics[width=0.2\linewidth]{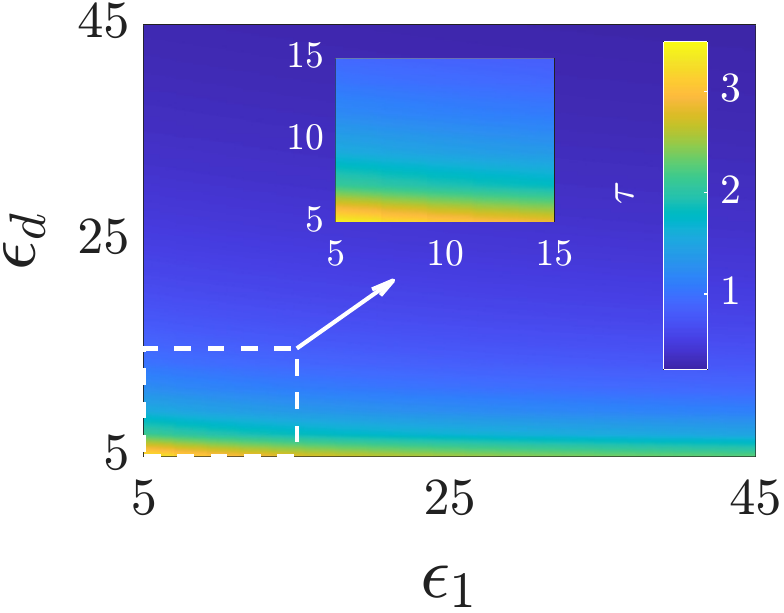}}
    
    \caption{(a)-(b): Variation of $\tau$ as a function of different values of $\epsilon_1$ and $\epsilon_d$, respectively. The theoretical variation of $\tau$, obtained from Eq.~\eqref{u-integral-e1-large}, is plotted in (a) using a solid green curve; (c)-(d): Variation of $r_{M}$ as a function of $t$, for $\epsilon_d=1$ and varying $\epsilon_1$, as well as for $\epsilon_1=6$ and varying $\epsilon_d$, respectively; \textcolor{red}{(e): Joint variation of $\tau$ (represented as a color-map) as a function of $(\epsilon_1,\epsilon_d)$, both varying within $[5,45]$, with the inset showing a zoomed view.} The circular markers in (a) and (b) represent different parameter combinations, whereas in (c) and (d), the horizontal and vertical lines have the same meaning as in Fig.~\ref{fig_r_ts_e1_6}. The interaction order is fixed at $d=2$ across all figures.}
    \label{vary_e1_ed}
\end{figure*}
The primary measure of transient dynamics considered in this paper is the FST, denoted by $\tau$, which is defined as the time taken by the order parameter $r(t)$, starting at $r(0)=r_0$, to reach a neighborhood $\delta\ll 1$ of the steady-state $r(t)=r_{\rm ss}$ such that $r(\tau)=r_{\rm ss}-\delta$; see Fig.~\ref{demo}c for a schematic. Clearly, this definition applies only to scenarios for which the order parameter reaches a stationary value, and the system parameters in the subsequent sections are chosen to reflect this limitation.
Following the steps in Appendix \ref{dervi_tau} and substituting $r^2=u$, $\tau$ can be expressed as
\begin{align}
    \tau&=\int_{u_0}^{u_{\rm ss}}\frac{\D u}{-2u\Delta + u(1-u)\left(\epsilon_1 + \epsilon_d u^{(d-1)}\right)}.
\label{u-integral}
\end{align}
For a system initialized near the asynchronous state, the lower limit $u_0=r_0^2\rightarrow 0^{+}$, whereas the upper limit $u_{\rm ss}=(r_{\rm ss}-\delta)^2$.

From Eq.~\eqref{u-integral}, two limiting cases arise. First, for large $\epsilon_1$ (\ie, $\epsilon_1\gg \epsilon_1^c=2\Delta,\epsilon_d$, such that $r_{\rm ss}\rightarrow 1$), Eq.~\eqref{u-integral} can be approximated as
\begin{equation}
    \tau\approx \frac{1}{\epsilon_1}\int_{u_0}^{(1-\delta)^2}\frac{\D u}{u(1-u)}=\underbrace{\ln\left(\frac{(1-\delta)^2 (1-u_0)}{u_0 [1-(1-\delta)^2]}\right)}_{C_0}\frac{1}{\epsilon_1},
\label{u-integral-e1-large}
\end{equation}
where $C_0$ is a constant dependent on only $u_0$ and $\delta$; thus, Eq.~\eqref{u-integral-e1-large} provides a theoretical estimate for $\tau$ as a function of $\epsilon_1$, albeit in a specific parameter range, and is independent of $(d,\epsilon_d)$. On the other hand, for large values of $d$, $u^{(d-1)}\approx 0$ when $u_{\rm ss}$ is sufficiently smaller than $1$. In this scenario, Eq.~\eqref{u-integral} can be approximated as
\begin{equation}
    \tau \approx \int_{u_0}^{u_{\rm ss}}\frac{\D u}{-2 u\Delta + \epsilon_1 u(1-u)},
\label{integral_large_d}
\end{equation}
thereby recovering pairwise-like dynamics in this regime.

\section{Numerical Results}
\label{num}

For all subsequent sections, $\Delta=1$, which implies $\epsilon_1^c=2$. In subsequent sections, the values of $(\epsilon_1,\epsilon_d)$ are chosen such that the system is constrained to a monostable steady synchronous regime (see Appendix \ref{phase_t}), in addition to $N=10^5$. This is because the FST is undefined in the case of non-stationary dynamics, which can arise either due to the stability of the order parameter dynamics or the effects arising from finite-sized oscillator ensembles. Furthermore, for all subsequent calculations and simulations, the parameters $r_0=\delta = 10^{-2}$. While arbitrary, this fixes the overall synchronization timescale, while preserving the qualitative observations reported subsequently; see Appendices \ref{init_choice} and \ref{delta_choice}.

\begin{figure*}[t]
    \centering
    \subfloat[\label{d_vary_e1_6_ed_5_a}]{\includegraphics[width=0.195\linewidth]{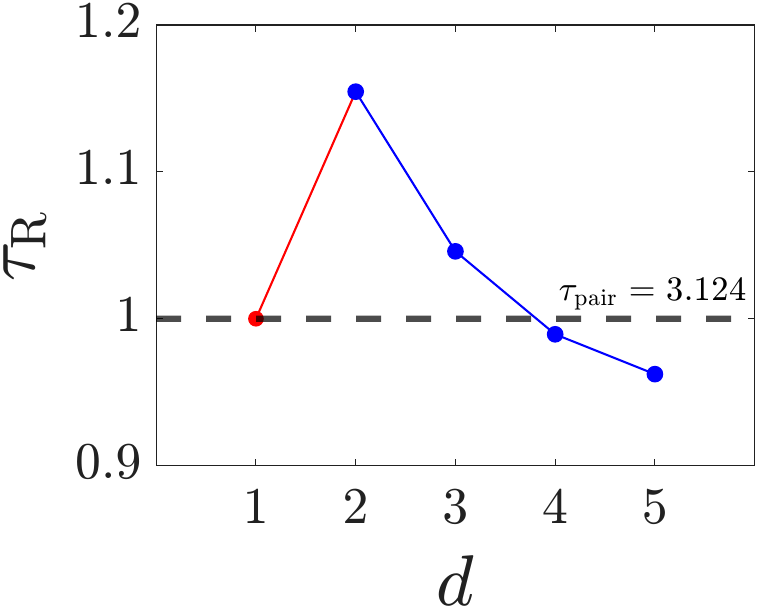}}
    \subfloat[$d=2$\label{d_vary_e1_6_ed_5_b}]{\includegraphics[width=0.2\linewidth]{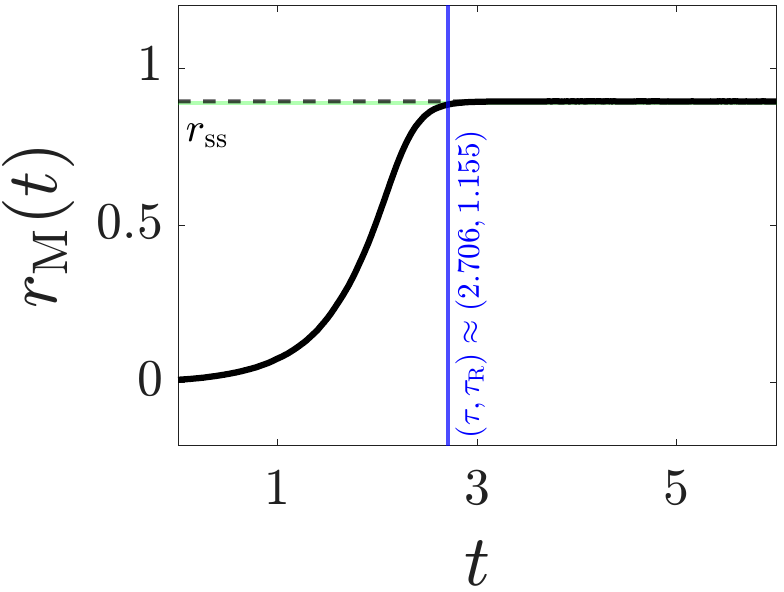}}
    \subfloat[$d=3$\label{d_vary_e1_6_ed_5_c}]{\includegraphics[width=0.2\linewidth]{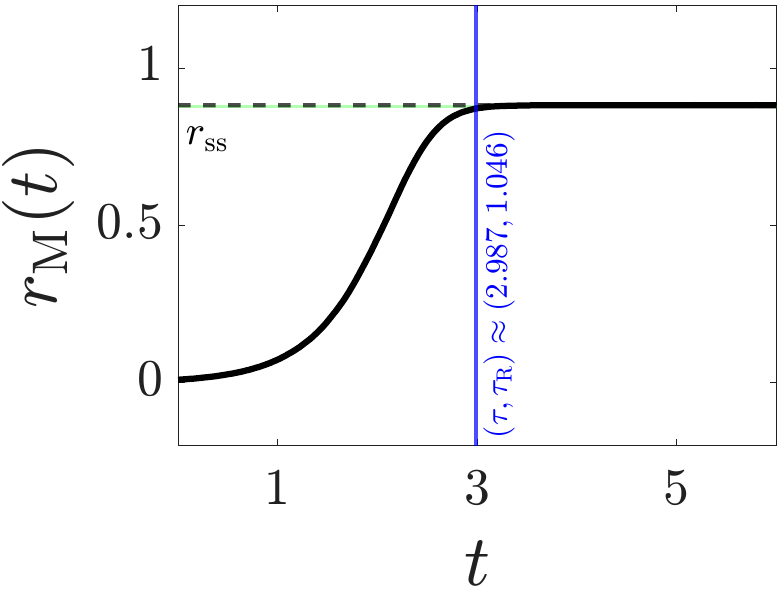}}
    \subfloat[$d=4$\label{d_vary_e1_6_ed_5_d}]{\includegraphics[width=0.2\linewidth]{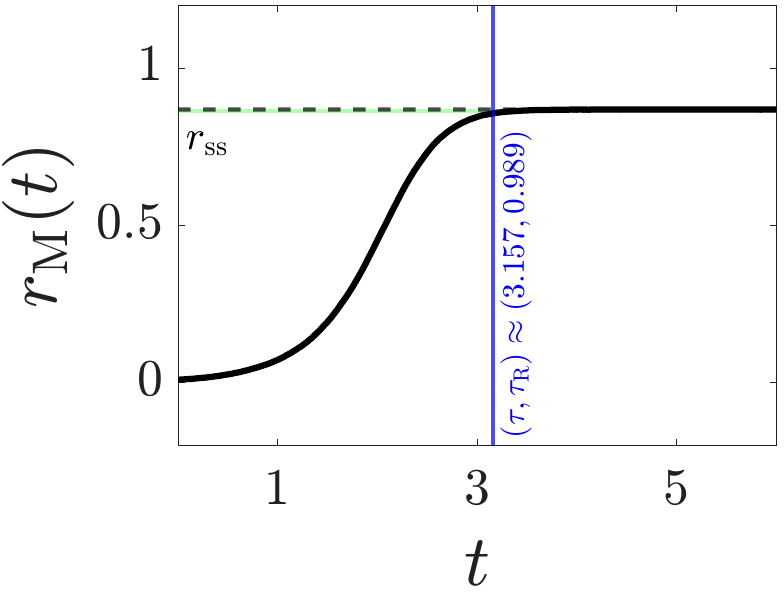}}
    \subfloat[$d=5$\label{d_vary_e1_6_ed_5_e}]{\includegraphics[width=0.2\linewidth]{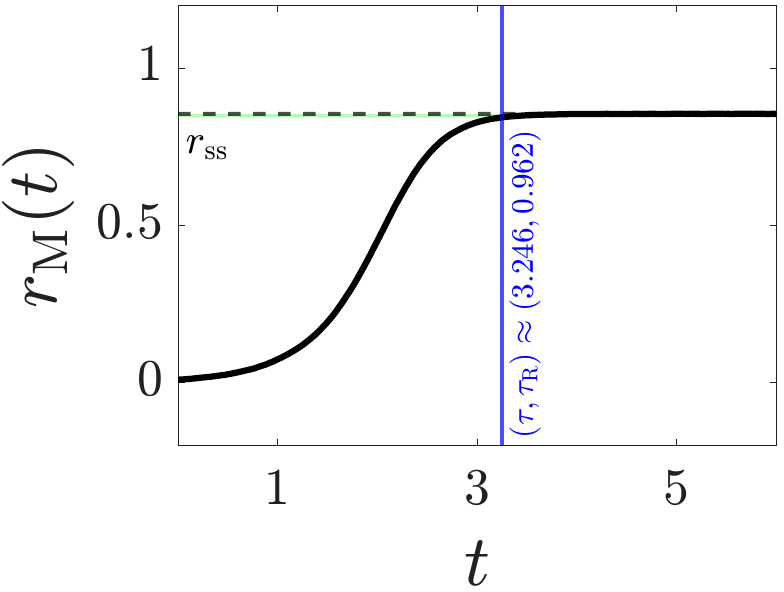}}

    \subfloat[$\epsilon_1=6$\label{d_vary_e1_6_9_12_15_18a}]{\includegraphics[width=0.2\linewidth]{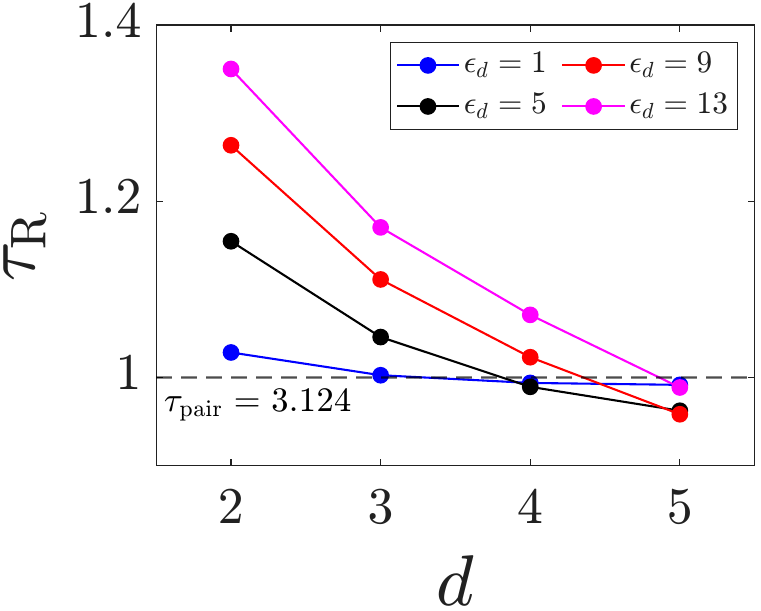}}
    \subfloat[$\epsilon_1=9$\label{d_vary_e1_6_9_12_15_18b}]{\includegraphics[width=0.2\linewidth]{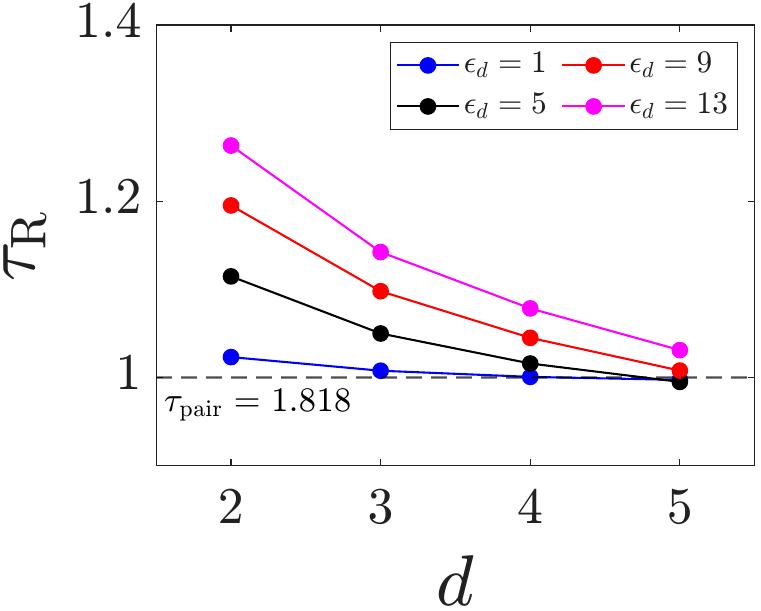}}
    \subfloat[$\epsilon_1=12$\label{d_vary_e1_6_9_12_15_18c}]{\includegraphics[width=0.2\linewidth]{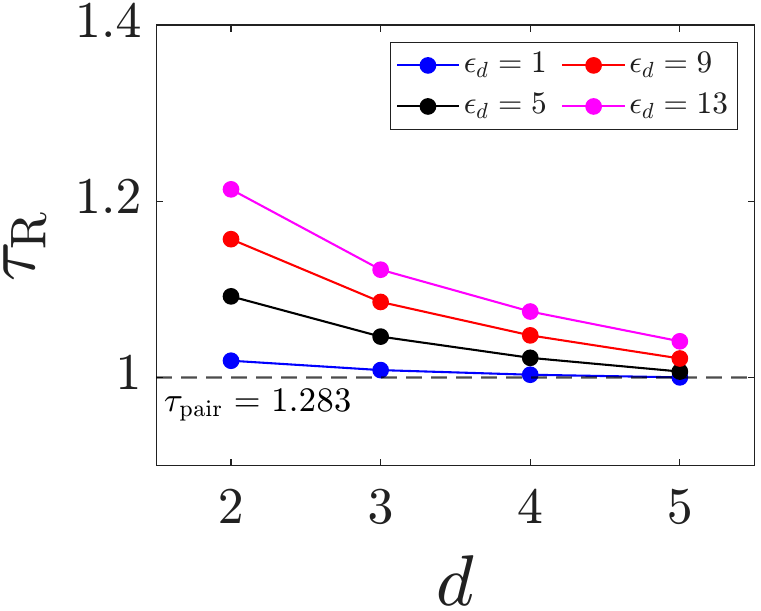}}
    \subfloat[$\epsilon_1=15$\label{d_vary_e1_6_9_12_15_18d}]{\includegraphics[width=0.2\linewidth]{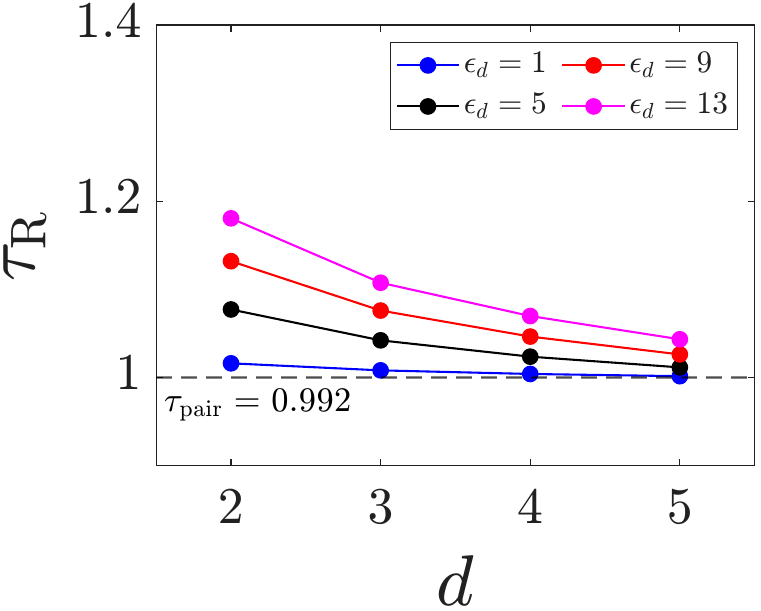}}
    \subfloat[$\epsilon_1=18$\label{d_vary_e1_6_9_12_15_18e}]{\includegraphics[width=0.2\linewidth]{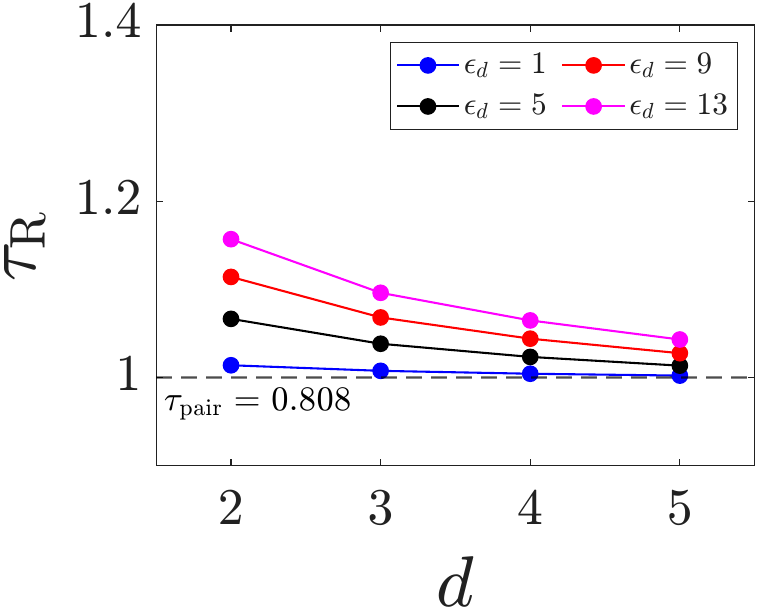}}
    \caption{\textcolor{red}{(a): Variation of $\tau_{\rm R}$ as a function of $d\in[1,5]$ for $(\epsilon_1,\epsilon_d)=(6,5)$. Here, the section of the curve in red highlights the change going from pairwise ($\tau_{\rm pair}=3.124$; marked with a broken horizontal line) to triadic interactions, whereas the remaining curve (in blue) captures the effects of further increasing the interaction order; (b)-(e): Variation of $r_{\rm M}$ as a function of $t$ for different values of $d$, whereas $(\epsilon_1,\epsilon_d)=(6,5)$. In these figures, the horizontal and vertical lines have the same meaning as in Fig.~\ref{fig_r_ts_e1_6}; (f)-(j): Variation of $\tau_{\rm R}$ as a function of $d\in[2,5]$, for different values of both $(\epsilon_1,\epsilon_d)$, the former being highlighted with different colors.}}
    \label{d_vary_e1_6_ed_5}
\end{figure*}

\subsection{Validation of Eq.~\eqref{u-integral}}
\label{validation}

The curves in Fig.~\ref{fig_r_ts_e1_6a}-\ref{fig_r_ts_e1_6d} presents multiple realizations of the order-parameter dynamics $r(t)$ (in gray), the corresponding median trajectory $r_{\rm M}(t)$ (in black) calculated over $100$ realizations, and the reduced Ott–Antonsen dynamics $r_{\rm OA}(t)$ (in red) obtained by numerically solving Eq.~\eqref{r-eqn}. The latter is depicted using a dashed line to demonstrate the nearly overlapping $r_{\rm OA}(t)$ and $r_{\rm M}(t)$ dynamics. The four sub-figures correspond to distinct parameter sets, where Fig.~\ref{fig_r_ts_e1_6a} includes only pairwise interactions, while Figs.~\ref{fig_r_ts_e1_6b}–\ref{fig_r_ts_e1_6d} incorporate both pairwise and higher-order interactions. The observations indicate that the individual trajectories reach within a neighborhood $\delta$ around $r_{\rm ss}$ at slightly different time instances due to differences in the initial realization of the phases and finite system size. However, the reduced and median dynamics are near-identical, with the value of $\tau$ accurately predicting the FST of the median trajectory (see insets) across qualitatively different parameter sets, thereby validating the expression in Eq.~\eqref{u-integral}.

\subsection{Effect of varying $(\epsilon_1, \epsilon_d)$, and $d$}
\label{var-results}

First, the subplots in Fig.~\ref{vary_e1_ed} demonstrate the effects of varying $\epsilon_1$ and $\epsilon_d$ in a system with pairwise and triadic interactions (\ie, $d=2$). In particular, Fig.~\ref{e1_vary} shows that, for a fixed $\epsilon_d$, $\tau$ decreases monotonically with increasing $\epsilon_1$ and exhibits an $\epsilon_1^{-1}$ scaling for large $\epsilon_1$, as predicted by Eq.~\eqref{u-integral-e1-large}. Similarly, Fig.~\ref{ed_vary} highlights that, while for lower values of $\epsilon_1$ (here, $\epsilon_1=4,6$), the value of $\tau$ decreases monotonically with increasing $\epsilon_d$, for relatively high values of $\epsilon_1$ (here, $\epsilon_1=8,16)$, the value of $\tau$ remains essentially constant as $\epsilon_d$ is increased.
These trends are corroborated via time series plots of $r_{\rm M}(t)$ in Figs.~\ref{rts_ed_1_d_2} and \ref{rts_e1_6_d_2}, where the FST, estimated both from Eq.~\eqref{u-integral} and from the median dynamics, is observed to decrease with increasing either $\epsilon_1$ or $\epsilon_d$, respectively, while keeping other parameters constant. \textcolor{red}{Additionally, a combined variation of both $(\epsilon_1,\epsilon_d)$ is presented in Fig.~\ref{param_plot_d2}, with the resulting value of $\tau$ plotted as a color map. The smooth gradient from yellow to blue indicates that the value of $\tau$ decreases monotonically (also see inset), thereby confirming the validity of the previous observations across a much larger parameter space\footnote{The lower bounds of $(\epsilon_1,\epsilon_d)$ are taken such that it avoids parameter values which might result in non-steady dynamics for $d=2$}.}

Next, to quantify the effect of higher-order interactions on FST, as compared to purely pairwise interactions, a measure $\tau_{\rm R}=\tau_{\rm pair}/\tau_{\rm d}$ is computed, where $\tau_{\rm d}$ and $\tau_{\rm pair}$ are the FSTs corresponding to $\epsilon_d\neq 0$ and $\epsilon_d=0$ (\ie, purely pairwise), respectively. Subsequently, Fig.~\ref{d_vary_e1_6_ed_5} demonstrates the variation of the FST as a function of $d\in[1,5]$ for $(\epsilon_1,\epsilon_d)=(6,5)$. In particular, the observations in Fig.~\ref{d_vary_e1_6_ed_5_a} suggest that increasing the order of interaction from pairwise to triadic leads to an increase in $\tau_{\rm R}$ (marked in red), followed by a monotonic reduction, which subsequently drops below the pairwise level (marked by $\tau_{\rm R}=1$) for $d=4,5$. These observations are further corroborated using the time series plots of $r_{\rm M}(t)$ in Figs.~\ref{d_vary_e1_6_ed_5_b}-\ref{d_vary_e1_6_ed_5_e}, which show the relatively faster and slower approach to $r_{\rm ss}$ for $d=2,3$ and $d=4,5$, respectively. \textcolor{red}{Furthermore, the monotonic decrease of $\tau_{\rm R}$ for $d>2$ hold qualitatively across different combinations of $\epsilon_1$ and $\epsilon_d$. This is demonstrated in Figs.~\ref{d_vary_e1_6_9_12_15_18a}-\ref{d_vary_e1_6_9_12_15_18e} for different values of $\epsilon_1=6,9,12,15,18$, and $\epsilon_d=1,5,9,13$.}
\begin{figure}[b]
    \centering
    \subfloat[$\epsilon_d=5$\label{large_d_plot_a}]{\includegraphics[width=0.5\linewidth]{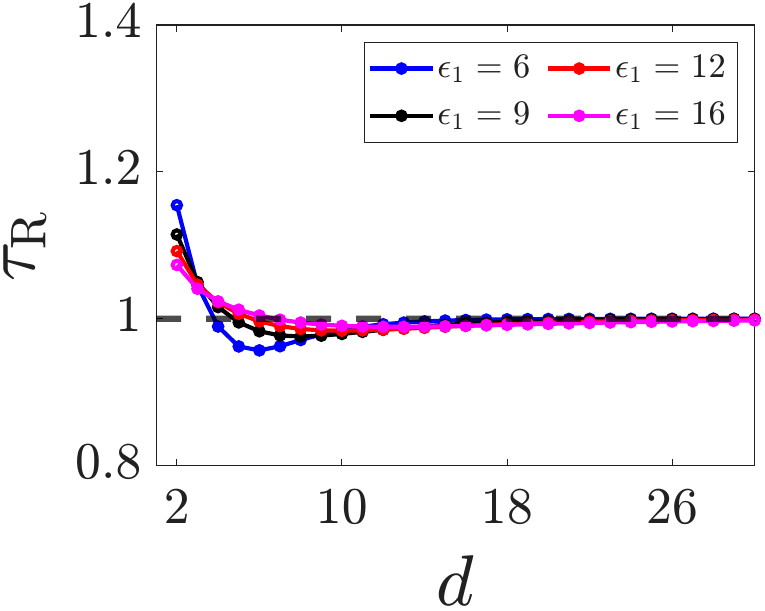}}\hfill
    \subfloat[$\epsilon_d=9$\label{large_d_plot_b}]{\includegraphics[width=0.5\linewidth]{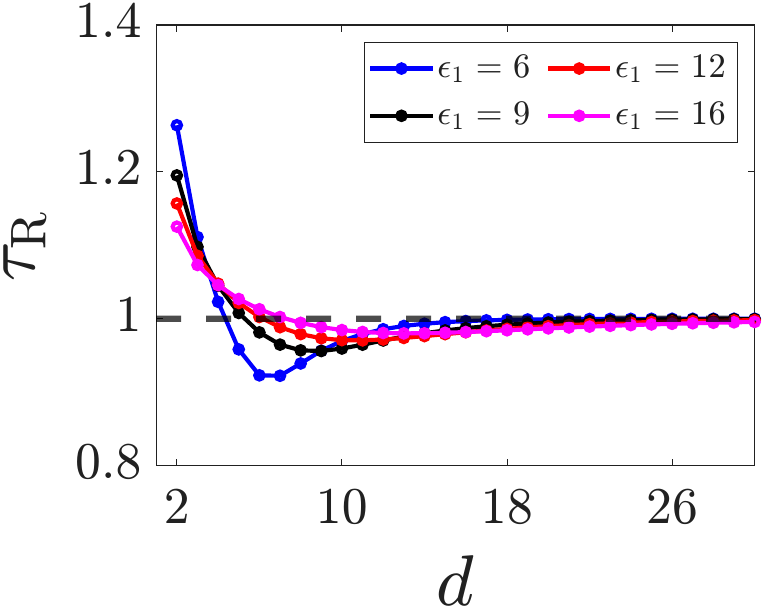}}

    \subfloat[\label{large_d_plot_c}]
    {\includegraphics[width=0.5\linewidth]{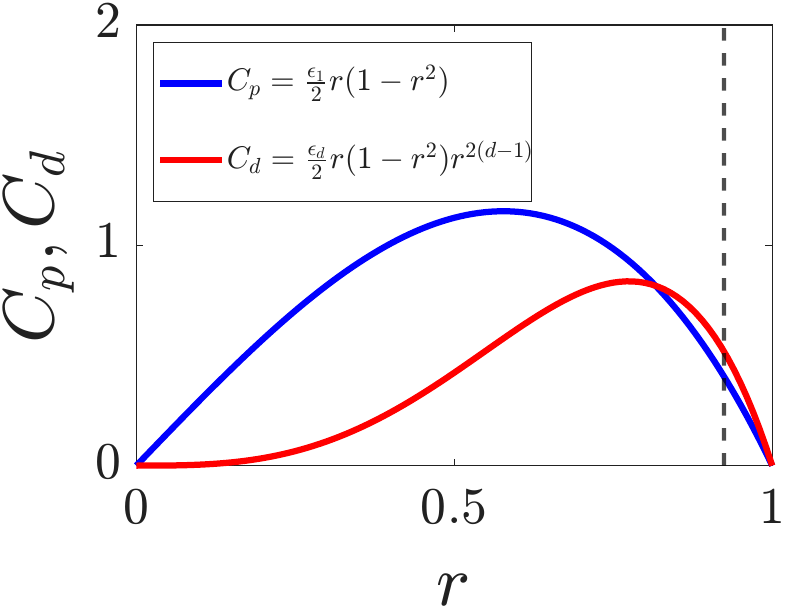}\includegraphics[width=0.5\linewidth]{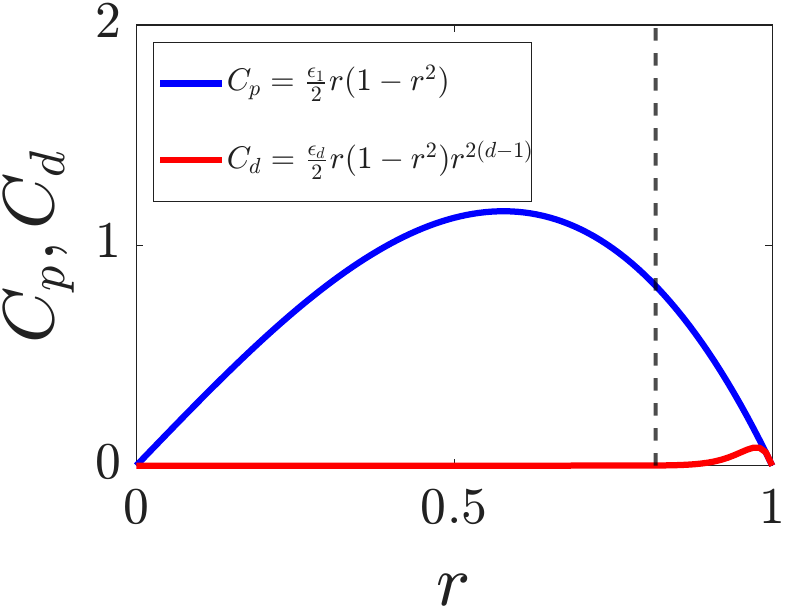}}

    \caption{(a)-(b): Variation of $\tau_{\rm R}$ as a function of $d\in[2,30]$ for $\epsilon_d=5$ and $\epsilon_d=9$, respectively. The different colors correspond to different values of $\epsilon_1$, whereas the horizontal line denotes $\tau_{\rm R}=1$; (c): Variation of $C_p$ (blue) and $C_d$ (red), as a function of $r$, for $d=2$ (left) and $d=20$ (right), respectively, whereas the vertical line denotes $r_{\rm ss}$.}
    \label{large_d_plot}
\end{figure}

Finally, for $d\gg 5$, the system dynamics reverts to pairwise-like dynamics, as indicated by Eq.~\eqref{integral_large_d}. This is verified in Figs.~\ref{large_d_plot_a} and \ref{large_d_plot_b} for different values of $\epsilon_1$ (marked in different colors), whereas $\epsilon_d=5,9$, respectively. It is observed that $\tau$ converges to $\tau_{\rm pair}$ (\ie, $\tau_{\rm R}\approx 1$), as $d$ increases beyond a certain range. However, these cannot be corroborated further using plots of $r_{\rm M}(t)$, since the necessary simulations become prohibitively expensive for interactions of such high order. Heuristically, this can be understood from Eq.~\eqref{r-eqn}, where, during most of the transient evolution, the order parameter $r(t)$ lies within the range $r(t) \ll r_{\rm ss}$. Since the higher-order coupling scales as $r^{2(d-1)}$, its contribution rapidly approaches zero for large values of $d$, whereas for smaller values of $d$, it can contribute significantly. This is demonstrated in Fig.~\ref{large_d_plot_c} for the case of $(\epsilon_1,\epsilon_d)=(6,9)$ for two different values of $d$. The curves $C_p$ (blue) and $C_d$ (red) denote the magnitude of the pairwise and higher-order terms (from Eq.~\eqref{r-eqn}), respectively, as a function of $r$. Clearly, for $d=2$, $C_d(r\leq r_{\rm ss})$ is sufficiently greater than $0$, whereas for $d=20$, $C_d(r\leq r_{\rm ss})\approx 0$. While this explains the asymptotic behavior and why triadic interactions cause faster transitions, the intermediate variation depends on the complex interplay between $C_p$ and $C_d$, the dynamical outcomes of which differ based on the system parameters.

\section{Summary \& Discussions}
\label{conc}
Working in the framework of a globally coupled and \textcolor{red}{deterministic} Kuramoto model, this paper investigates the effects of higher-order interactions on synchronization dynamics, with a particular focus on quantifying the transient phase. By assuming an infinitely large ensemble of oscillators having Cauchy-distributed natural frequencies, the reduced dynamics are obtained using the Ott-Antonsen ansatz. The resulting differential equation is then integrated over suitable limits to obtain an expression for the FST. Considering its complicated form, it is subsequently evaluated numerically for parameter values in the monostable synchronous regime and shown to capture how quickly the median dynamics, computed over many independent realizations of the system equations, converge to the steady state.

The obtained results highlight that, for triadic interactions, increasing the magnitude of either pairwise or higher-order coupling strength leads to a faster approach to the steady state. In contrast, increasing the interaction order beyond the pairwise case, while keeping the other system parameters fixed, yields nontrivial behavior. Specifically, increasing from pairwise to triadic leads to a faster transition; however, further increases to tetradic, pentadic, and hexadic orders induce progressively slower transitions, which may even fall below the pairwise case for certain parameter combinations. Additionally, it is also observed that, for very large orders of interaction, the transition time returns to the pairwise level. This can be attributed to the fact that, for higher-order interactions to contribute, the system must already be substantially synchronized. However, this is not achieved in the transient phase, and therefore, the system is effectively governed by only the pairwise couplings. 
Taken together, these findings provide fundamental insights into the dynamics of complex systems.

However, due to the challenges associated with higher-order interactions, the present study is confined to a small parametric regime. \textcolor{red}{In contrast, realistic complex systems are often influenced by additional effects such as stochastic fluctuations arising from unresolved physical mechanisms, heterogeneous and time-varying network connectivity, non-standard system parameters and natural frequency distributions, as well as finite-size effects. These factors can substantially impact the transients and the corresponding FST, and become particularly relevant in models describing neural dynamics, climatic events, and propagation of contagions. While these have seen some recent contributions (for example, Zhao \etal~\cite{zhao2026synchronization} and Ham \etal~\cite{ham2024stochastic}), systematically addressing these limitations in the context of the FST is an open area with several important directions for possible future research}. Additionally, investigating how interaction order influences other emergent phenomena (for example, swarming~\cite{o2017oscillators}), extending the FST to non-stationary regimes, and examining how higher-order interactions may be utilized to regulate the transient state of engineered systems are interesting questions that can be explored further.

\begin{acknowledgments}
DB and AB acknowledge financial support from NBRC Gurgaon Core Funds, and DB also acknowledges financial support provided by IIT Kharagpur, as well as discussions with Dr. Proloy Das (NBRC Gurgaon) and Dr. Somnath Roy (IEM Kolkata). AB was supported by the Dementia Science Program, Department of Biotechnology, Government of India, and  Ministry of Youth Affairs and Sports (MYAS), Government of India (F.NO.K-15015/42/2018/SP-V).
\end{acknowledgments}

\section*{Author Declarations}
\subsection*{Conflict of Interest}
The authors have no conflicts of interest to disclose.

\subsection*{Author Contributions}

\textbf{DB}: Conceptualization, Data Curation, Formal Analysis, Investigation, Methodology, Software, Validation, Visualization, Writing/Original Draft Preparation. \textbf{PP}: Investigation, Funding Acquisition, Project Administration, Supervision, Writing/Review \& Editing. \textbf{AB}: Conceptualization, Investigation, Visualization, Funding Acquisition, Project Administration, Supervision, Writing/Review \& Editing.

\section*{Data Availability Statement}
The data that support the findings of this study are available from the corresponding author upon reasonable request.

\appendix

\section{Deriving the order parameter dynamics}
\label{dervi_tau}

Assuming $N$ to be large and a Cauchy distribution of natural frequencies having zero mean and scale factor $\Delta$, the dynamics of the complex Kuramoto order parameter $z(t)$, defined as
\begin{equation}
    z(t)=r(t)e^{\img \psi(t)}=\frac{1}{N}\sum_{i=1}^N e^{\img \phi_i(t)},\ \img=\sqrt{-1},
\end{equation}
can be expressed~\cite{ott2008low,ott2009long,biswas2026emergent} in terms of $r(t)$ and $\psi(t)$ as
\begin{align}
    \frac{\D r(t)}{\D t}=R(r)=- r\Delta + r(1-r^2)\left[\frac{\epsilon_1}{2}+\frac{\epsilon_d}{2}r^{2(d-1)}\right],
\label{r-eqn}
\end{align}
and $\dot{\psi}(t)=0$. Subsequently, the steady-state solution of Eq.~\eqref{r-eqn} (say, $r_{\rm ss}$) can be obtained as a solution of the algebraic equation
\begin{equation}
    \left(\frac{2\Delta}{1-r^2_{\rm ss}}\right)=\epsilon_1+\epsilon_d r_{\rm ss}^{2(d-1)}.
\label{r_ss_value}
\end{equation}
While the asynchronous state (\ie, $r(t)=0$) is stable for all $\epsilon_1\leq \epsilon_1^c= 2\Delta$ (\ie, the forward transition point), the stability of $r_{\rm ss}$ can be obtained from $\D R/\D r|_{r_{\rm ss}}\leq 0$, which translates to 
\begin{equation}
    \epsilon_d(d-1)(1-r_{\rm ss}^2)^2 r_{\rm ss}^{2(d-2)}\leq 2\Delta.
\end{equation}
Given its complicated structure, a closed form solution of Eq.~\eqref{r-eqn} can be obtained for the purely pairwise case, \ie, $\epsilon_d=0$, as
\begin{equation}
    r(t)=r_{\rm ss}\left[1-\left(1-\left[\frac{r_{\rm ss}}{r_0}\right]^2\right)e^{(2\Delta-\epsilon_1)t}\right]^{-1/2},
\label{r_sol_d1}
\end{equation}
where $r_0\equiv r(0)$ denotes the initial value of the order-parameter and $r^2_{\rm ss}=1-(2\Delta/\epsilon_1)$.
However, for $\epsilon_{d}\neq 0$, Eq.~\eqref{r-eqn} can only be integrated numerically. 

\begin{figure}[t]
    \centering
    
    \subfloat[$d=2$, $\epsilon_d=1$]{\includegraphics[width=0.327\linewidth]{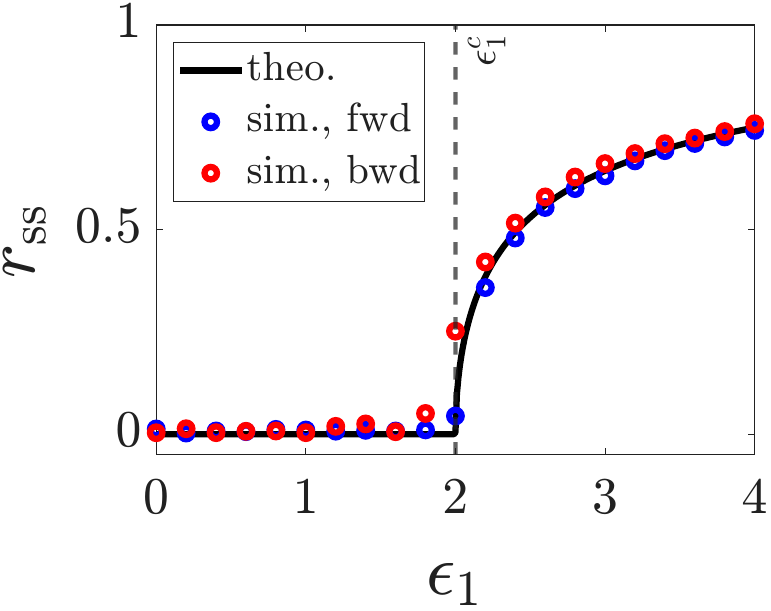}}\hfill
    \subfloat[$d=2$, $\epsilon_d=5$]{\includegraphics[width=0.327\linewidth]{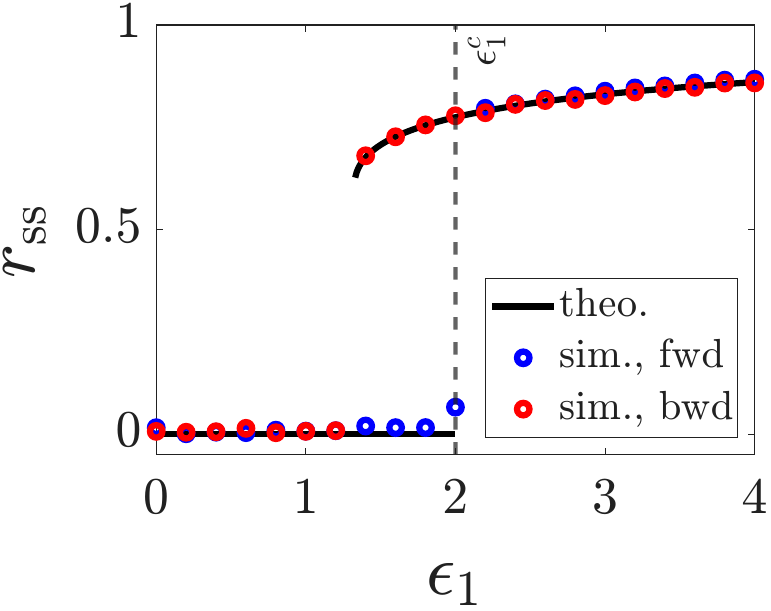}}\hfill
    \subfloat[$d=2$, $\epsilon_d=9$]{\includegraphics[width=0.327\linewidth]{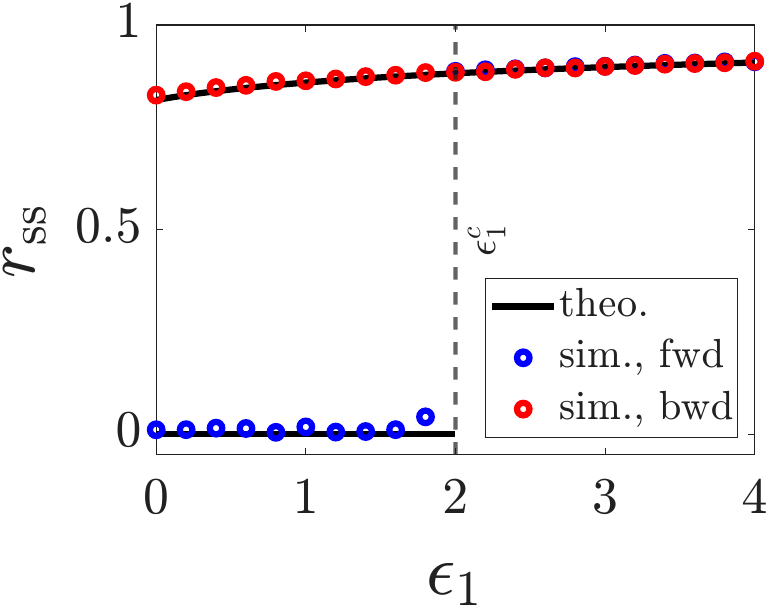}}

    \subfloat[$d=3$, $\epsilon_d=1$]{\includegraphics[width=0.327\linewidth]{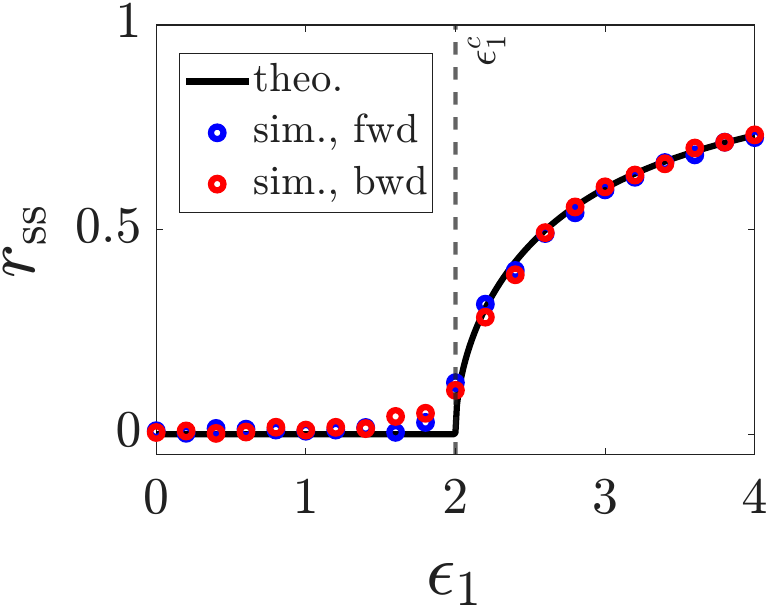}}\hfill
    \subfloat[$d=3$, $\epsilon_d=5$]{\includegraphics[width=0.327\linewidth]{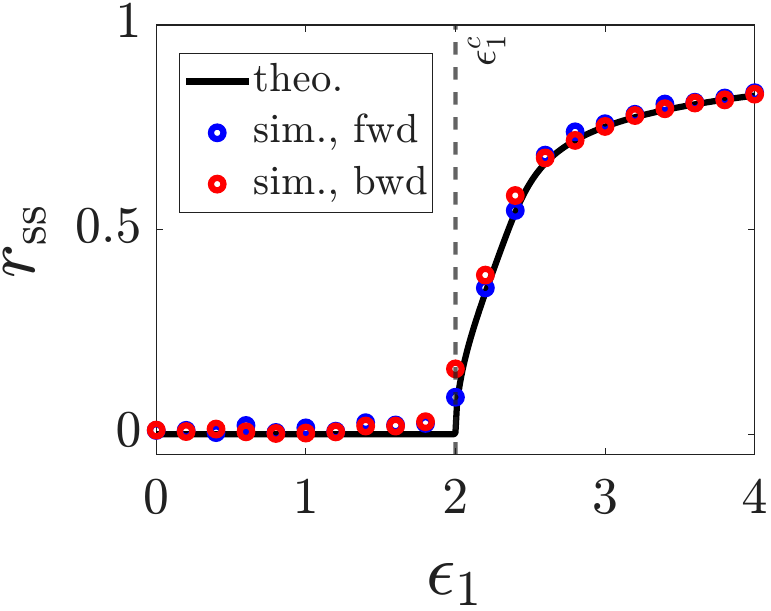}}\hfill
    \subfloat[$d=3$, $\epsilon_d=9$]{\includegraphics[width=0.327\linewidth]{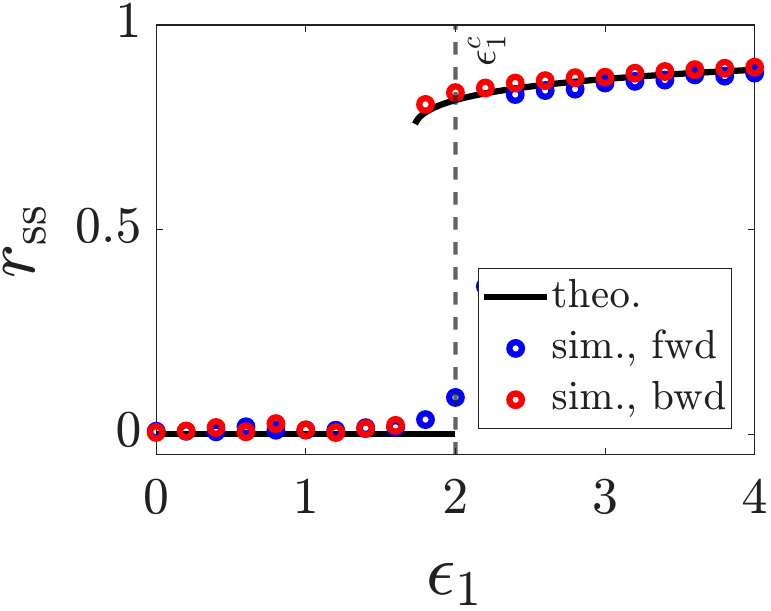}}

    \subfloat[$d=4$, $\epsilon_d=1$]{\includegraphics[width=0.327\linewidth]{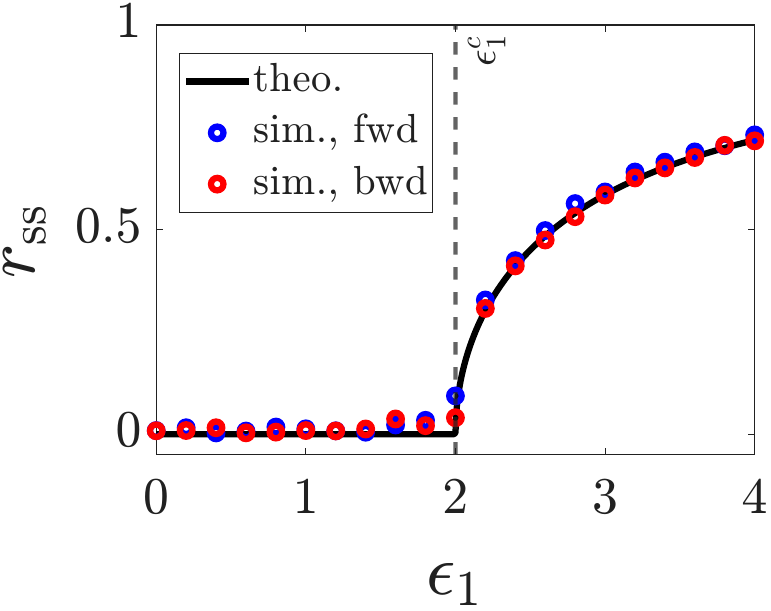}}\hfill
    \subfloat[$d=4$, $\epsilon_d=5$]{\includegraphics[width=0.327\linewidth]{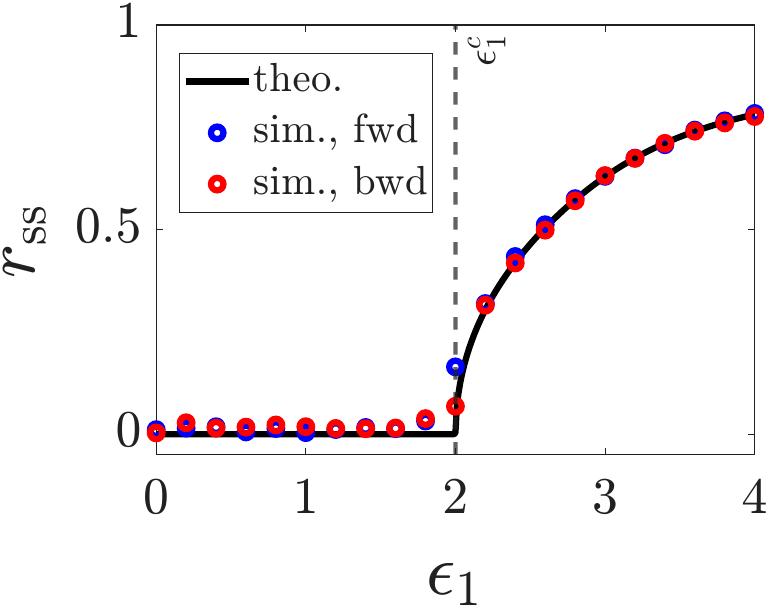}}\hfill
    \subfloat[$d=4$, $\epsilon_d=9$]{\includegraphics[width=0.327\linewidth]{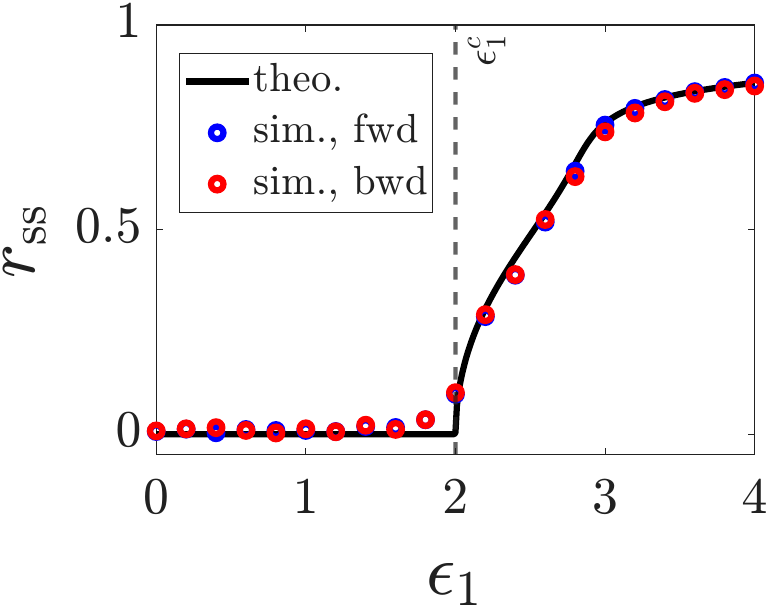}}

    \caption{(a)-(i): Variation of $r_{\rm ss}$ as a function of $\epsilon_1$, for different values of $(d,\epsilon_d)$. In each figure, the solid black line denotes the theoretical estimate of $r_{\rm ss}$, which is either zero, $\forall\ \epsilon_1\leq\epsilon_1^c$, or obtained by solving Eq.~\eqref{r_ss_value}, $\forall\ \epsilon_1>\epsilon_1^c$. In contrast, the circular markers denote the value of $r_{\rm ss}$ obtained by simulating Eq.~\eqref{main-eqn}, with the blue and red colors indicating the forward and backward variation, respectively, of $\epsilon_1$.}
    \label{fig_pt}
\end{figure}

\section{Synchronization phase transitions}
\label{phase_t}

The curves in Fig.~\ref{fig_pt} show a few representative examples of transitions observed in a system governed by equations of the form Eq.~\eqref{main-eqn}. The sub-figures highlight both continuous and explosive transitions, depending on the value of $d$ and $\epsilon_d$, characterized by either a smooth or sudden increase in the value of $r_{\rm ss}$, respectively, as the value of $\epsilon_1$ is tuned across the critical point. 

\section{Choice of $r_0$ and numerical integration}
\label{init_choice}

\begin{figure}[t]
    \centering
    \subfloat[\label{integrand_plots}]{\includegraphics[width=0.495\linewidth]{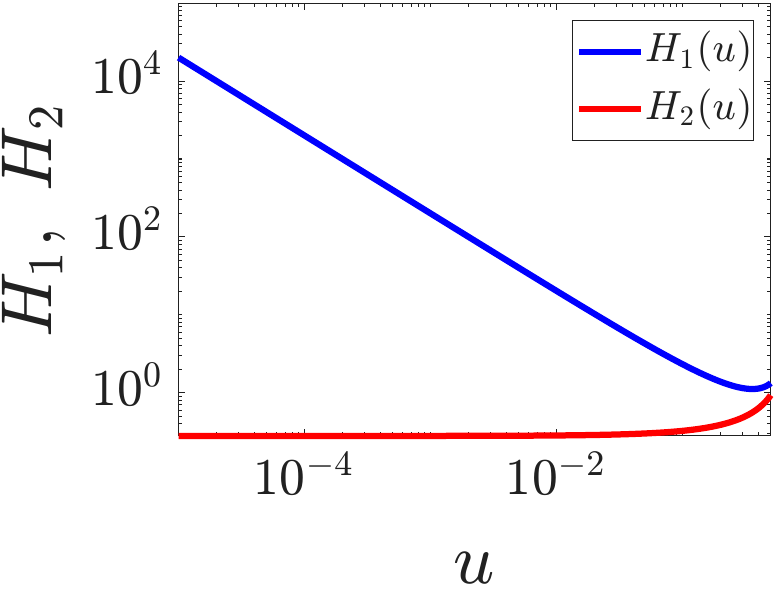}}\hfill
    \subfloat[\label{r0_vary}]{\includegraphics[width=0.495\linewidth]{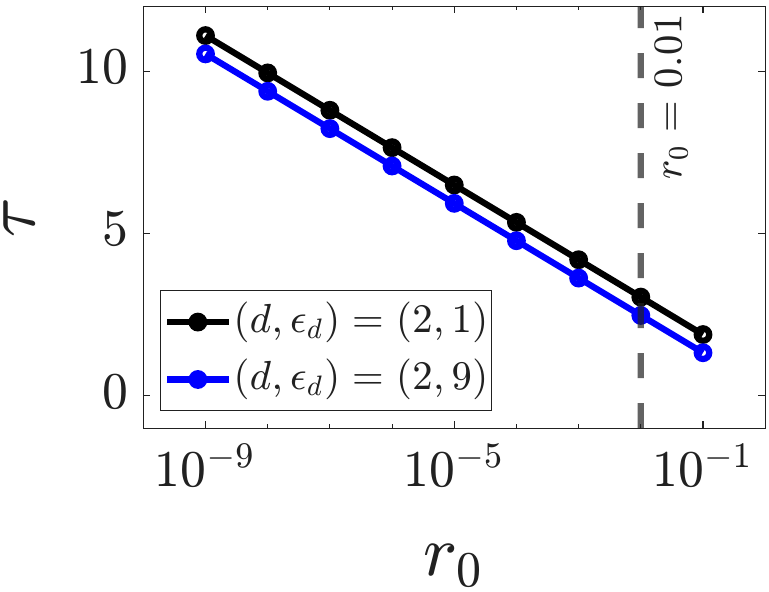}}

    \subfloat[$r_0=10^{-3}$\label{r0_m3}]{\includegraphics[width=0.495\linewidth]{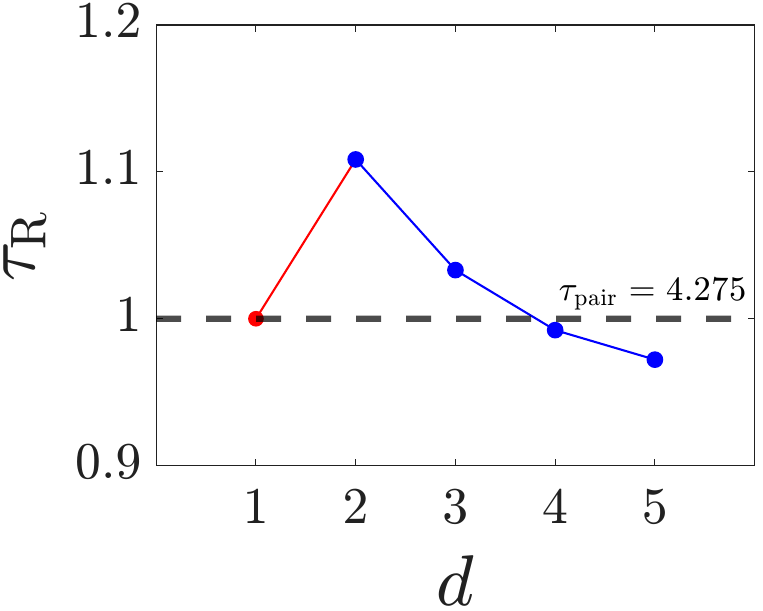}}\hfill
    \subfloat[$r_0=10^{-4}$\label{r0_m4}]{\includegraphics[width=0.495\linewidth]{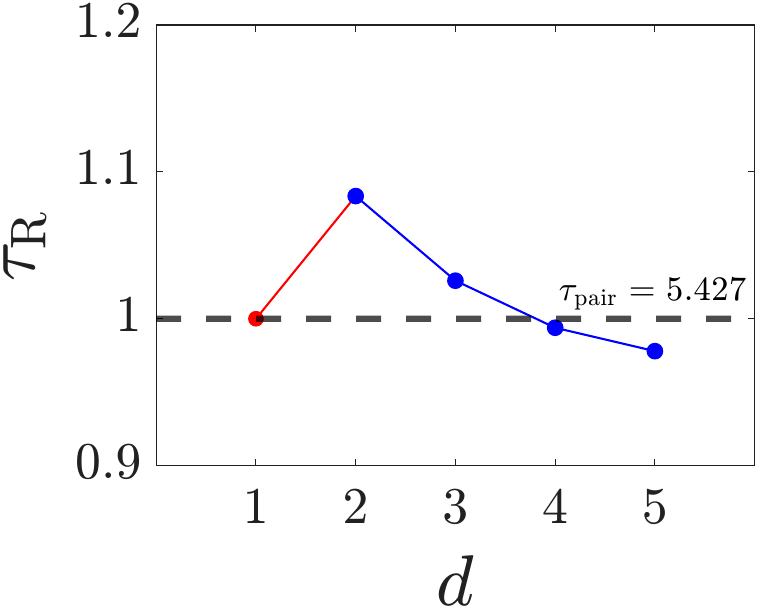}}
    \caption{(a) : Variation of $H_1$ (in blue) and $H_2$ (in red) as a function of $u$, demonstrating their behavior as $u\rightarrow 0$; (b): $\tau$ as a function of $r_0$ for $\epsilon_1 = 6$. Here, black and blue curves correspond to $(d,\epsilon_d) = (2,1)$ and $(2,9)$, respectively, whereas the vertical line denotes $r_0=0.01$, which is the assumption throughout the rest of the paper; (c)-(d): Variation of $\tau_{\rm R}$ as a function of $d$ (similar to Fig.~\ref{d_vary_e1_6_ed_5_a}), for $r_0=10^{-3}$ and $r_0=10^{-4}$, respectively.}
\end{figure}

The value of $\tau$ varies substantially with respect to $r_0^2$, and can be partly attributed to the singularity in the integrand. Therefore, to tackle this integral, Eq.~\eqref{u-integral} is rewritten as
\begin{align}
\begin{split}
    \tau&=\int_{u_0}^{u_{\rm ss}}\underbrace{\frac{1}{u\ G(u)}}_{H_1(u)}\D u \\ &=\underbrace{\int_{u_0}^{u_{\rm ss}}\frac{1}{u\ G(0)}\ \D u}_{\frac{1}{G(0)}\ln\left(u_{\rm ss}/u_0\right)} + \int_{u_0}^{u_{\rm ss}}\underbrace{\frac{1}{u}\left(\frac{1}{G(u)}-\frac{1}{G(0)}\right)}_{H_2(u)}\D u,\label{final_u_form}
\end{split}
\end{align}
where $G(u)=(1-u)(\epsilon_1 + \epsilon_d u ^{(d-1)})-2$ and $u=r^2$. Clearly, the first term can be integrated analytically, and it captures the effect of the divergence, which turns out to be logarithmic in nature. On the other hand, the second term can be integrated using simple numerical techniques since it does not diverge; see Fig.~\ref{integrand_plots}.
Subsequently, the curves in Fig.~\ref{r0_vary} demonstrate the variation of $\tau$ as a function of $r_0$, where it is observed that $\tau$ increases sharply with decreasing $r_0$. More importantly, although the absolute value of the FST depends sensitively on $r_0$, the qualitative non-monotonic dependence of the FST on the interaction order remains unchanged across different choices of $r_0$, as demonstrated in Figs.~\ref{r0_m3} and \ref{r0_m4}. Therefore, without loss of generality, $r_0=10^{-2}$ is chosen as the initial value of the order parameter dynamics in the paper.

\begin{figure}[b]
    \centering
    \subfloat[$d=2$\label{delta_vary_fig_a}]{\includegraphics[width=0.5\linewidth]{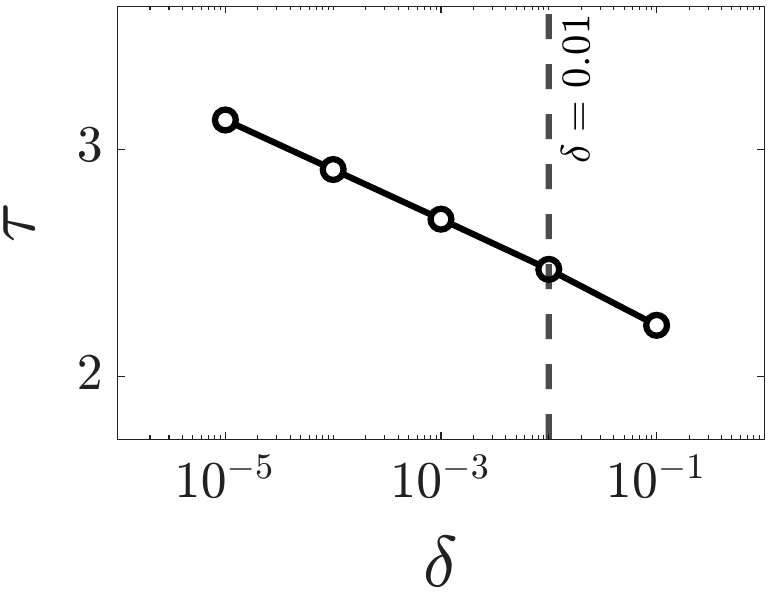}}
    \subfloat[$d=3$\label{delta_vary_fig_b}]{\includegraphics[width=0.5\linewidth]{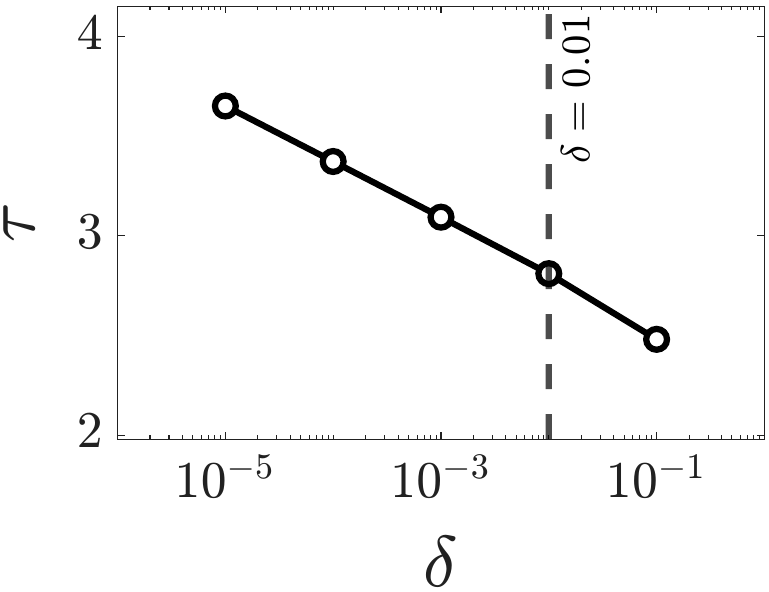}}
    
    \subfloat[$d=4$\label{delta_vary_fig_c}]{\includegraphics[width=0.5\linewidth]{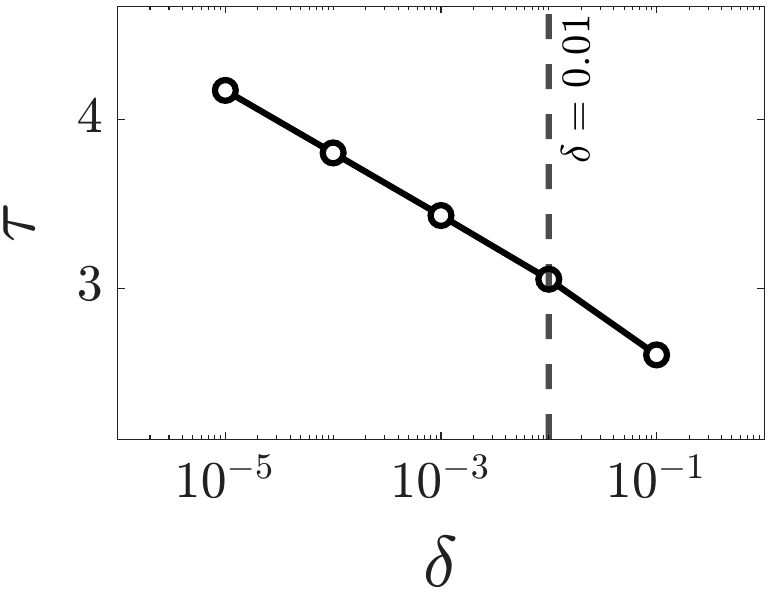}}
    \subfloat[$d=5$\label{delta_vary_fig_d}]{\includegraphics[width=0.5\linewidth]{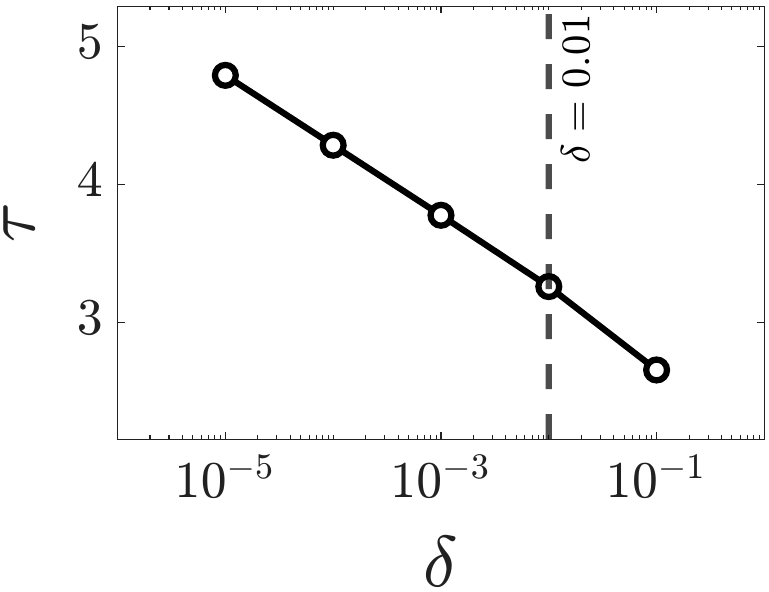}}
    \caption{(a)-(d): Variation of $\tau$ as a function of $\delta$ for $(\epsilon_1,\epsilon_d,r_0)=(6,9,10^{-2})$ and different values of $d$, whereas the broken vertical line denotes $\delta=0.01$, which is the assumption throughout the rest of the paper.}
    \label{delta_vary_fig}
\end{figure}

\section{Choice of $\delta$}
\label{delta_choice}

Akin to Appendix \ref{init_choice}, the value of $\tau$ is also dependent on the choice of the neighborhood parameter $\delta$. To get an analytical expression, first, Eq.~\eqref{r-eqn} is rearranged as
\begin{equation}
    \tau=\int_{r_0}^{(r_{\rm ss}-\delta)}\frac{1}{R(r)}\D r.
\label{d-expression}
\end{equation}
To evaluate this, the RHS is first expanded as
\begin{equation}
    \int_{r_0}^{(r_{\rm ss}-\delta)}\frac{1}{R(r)}\D r=\int_{r_0}^{(r_{\rm ss}-\xi)}\frac{1}{R(r)}\D r+\int_{(r_{\rm ss}-\xi)}^{(r_{\rm ss}-\delta)}\frac{1}{R(r)}\D r.
\label{basic-expansion}
\end{equation}
For $\delta\rightarrow 0$ and $\delta\ll \xi \ll 1$, the second term in the integral is expanded in a Taylor series around $r=r_{\rm ss}$. Ignoring second-order contributions, Eq.~\eqref{d-expression} can be approximated as
\begin{equation}
    \tau\sim \int_{(r_{\rm ss}-\xi)}^{(r_{\rm ss}-\delta)}\frac{1}{(r_{\rm ss}-r)}\D r,
\end{equation}
where some constants are suppressed for the sake of clarity. This clearly suggests that $\tau\sim-\ln(\delta)$ in the limiting case of $\delta\rightarrow 0$, up to some additive and multiplicative constants. This is verified in Figs.~\ref{delta_vary_fig_a}-\ref{delta_vary_fig_d} by numerically integrating the expression for $\tau$ for values of $\delta\in[10^{-5},0.5]$. It is observed that the curves form a straight line with a negative slope in a semi-log plot, similar to that in Fig.~\ref{r0_vary}. This suggests, similar to $r_0$, the value of $\delta$ also sets the overall timescale and does not affect the qualitative behavior of the transient dynamics. Therefore, without loss of generality, $\delta=10^{-2}$ is chosen as the neighborhood parameter in this article.

\bibliographystyle{unsrt}
\bibliography{refs}

\end{document}